\documentclass[english,12pt]{article}
\usepackage{array}
\usepackage{graphicx}
\usepackage{amssymb}
\usepackage{amsmath}
\usepackage{multirow}
\usepackage{prettyref}
\usepackage{babel}
\usepackage{units}
\usepackage[latin1]{inputenc}
\usepackage{amsfonts}
\usepackage{amssymb}
\usepackage{babel}
\usepackage{color}
\usepackage{cite}
\def\@fmsl@sh#1#2#3{\m@th\ooalign{$\hfil#1\mkern#2/\hfil$\crcr$#1#3$}}
 \def\eq#1\en{\begin{equation}#1\end{equation}}
\def\s[#1,#2]{[#1\stackrel{\star}{,}#2]}
\def\sx[#1,#2]{[#1\stackrel{\star_{x}}{,}#2]}

\newcommand{\nc}{\newcommand}
\nc{\beq}{\begin{equation}}
\nc{\eeq}{\end{equation}}
\nc{\beqa}{\begin{eqnarray}}
\nc{\eeqa}{\end{eqnarray}}

\def\bc{\begin{center}}
\def\ec{\end{center}}

\def\gsim{\mathrel{\mathpalette\atversim>}}

\def\bc{\begin{center}}
\def\ec{\end{center}}

\def\gsim{\mathrel{\rlap{\lower4pt\hbox{\hskip1pt$\sim$}}

    \raise1pt\hbox{$>$}}}       

\def\gsim{\mathrel{\rlap{\lower4pt\hbox{\hskip1pt$\sim$}}
    \raise1pt\hbox{$>$}}}       

\begin{document}
\makeatletter
\def\fmslash{\@ifnextchar[{\fmsl@sh}{\fmsl@sh[0mu]}}
\def\fmsl@sh[#1]#2{%
  \mathchoice
    {\@fmsl@sh\displaystyle{#1}{#2}}%
    {\@fmsl@sh\textstyle{#1}{#2}}%
    {\@fmsl@sh\scriptstyle{#1}{#2}}%
    {\@fmsl@sh\scriptscriptstyle{#1}{#2}}}
\def\@fmsl@sh#1#2#3{\m@th\ooalign{$\hfil#1\mkern#2/\hfil$\crcr$#1#3$}}
\makeatother

\thispagestyle{empty}
\begin{titlepage}
\boldmath
\begin{center}
  \Large {\bf Cosmological Evolution of the Higgs Boson's Vacuum Expectation Value}
    \end{center}
\unboldmath
\vspace{0.2cm}
\begin{center}
{  {\large Xavier Calmet}\footnote{x.calmet@sussex.ac.uk}} 
 \end{center}
\begin{center}
{\sl Department of Physics $\&$ Astronomy, 
University of Sussex, Brighton, BN1 9QH, United Kingdom 
}
\end{center}
\begin{abstract}
\noindent
We point out that the expansion of the universe leads to a cosmological time evolution of the vacuum expectation of the Higgs boson. Within the standard model of particle physics, the cosmological time evolution of the vacuum expectation of the Higgs leads to a cosmological time evolution of the masses of the fermions and of the electroweak gauge bosons while the scale of Quantum Chromodynamics  (QCD) remains constant. Precise measurements of the cosmological time evolution of $\mu=m_e/m_p$, where $m_e$ and $m_p$ are respectively the electron and proton mass (which is essentially determined by the QCD scale), therefore provide a test of the standard models of particle physics and of cosmology.  This ratio can be measured using modern atomic clocks.

\end{abstract}  
\end{titlepage}

The idea that physical constants could experience a  cosmological time evolution has received much attention, see for example \cite{Dirac,Milne,Jordan,Marciano:1983wy,Uzan:2002vq,Calmet:2001nu,Calmet:2002ja,Calmet:2006sc,Passarino:2001wx,Casadio:2007ip,Yoo:2002vw,Dent:2001ga,Calmet:2014qxa,Fritzsch:2016ewd,Sola:2015xga}. The most recent of these  investigations were motivated by cosmological observations that some of the fundamental constants of nature may not be that constant after all, see e.g. \cite{Webb:2000mn}.
In this work, we point out that the expansion of the universe leads to a cosmological time evolution of the vacuum expectation of the Higgs boson. Within the standard model of particle physics, the cosmological time evolution of the vacuum expectation of the Higgs boson leads to a cosmological time evolution of the masses of the fermions and of the electroweak gauge bosons while the scale of Quantum Chromodynamics (QCD) and thus the proton mass remain constant. Strictly speaking, quark masses also contribute to the proton mass and would lead to a small time dependence of the proton mass, but this is a tiny and thus negligible effect as the main contribution from the QCD scale to the proton mass remains constant. We show that precise measurements of the cosmological time evolution of $\mu=m_e/m_p$, where $m_e$ and $m_p$ are respectively the electron and proton mass, therefore provide a test of the standard models of particle physics and of cosmology.  

The discovery of the Higgs boson at the Large Hadron Collider in 2012 with a mass of 125 GeV was an amazing confirmation of the standard model of particle physics. The standard model of cosmology $\Lambda$CDM which posits the existence of cold dark matter and of a cosmological constant is equivalently successful. The cosmological model assumes that the expansion of the universe is described by the Friedmann-Lema\^itre-Robertson-Walker  (FLRW) metric
\begin{eqnarray}
ds^2=-dt^2+R^2(t) \left [ \frac{dr^2}{1-k r^2}+r^2 d\Omega^2\right]
\end{eqnarray}
which corresponds to an homogeneous and isotropic universe. Here $k$ is the curvature signature and $R$ is the expansion factor, whose time change is given by the Friedmann equation
\begin{eqnarray}
H(t)^2=\left ( \frac{\dot R}{R} \right)^2= \frac{8 \pi G}{3} \rho_{\mbox{tot}} -\frac{k}{R^2},
\end{eqnarray}
where the cosmological constant is included in the total energy density $\rho_{\mbox{tot}}$. The total energy density contains not only the dark energy but also dark matter and the visible matter. The value of the cosmological constant is such that our universe is currently undergoing a phase of accelerated expansion as the cosmological constant starts to dominate over all other forms of energy.

In the standard model of particle physics, the Higgs boson is part of a SU(2)$_L$ doublet ${\cal H}$ which in the unitarity gauge takes the form ${\cal H}=1/\sqrt{2}(0, \phi(\vec x,t)+v)^\top$ where $v=246$ GeV is the vacuum expectation value of the Higgs field $\phi(\vec x,t)$. In flat space-time $v$ is a constant, Lorentz invariant, quantity. We now show that the expansion of the universe leads to a cosmological time evolution of the vacuum expectation of the Higgs boson. We will show that within the standard model of particle physics, this cosmological time evolution of the vacuum expectation of the Higgs boson leads to a cosmological time evolution of the masses of the fermions and of the electroweak gauge bosons. 

The scalar sector of the standard model in the unitarity gauge is described by the following action
\begin{eqnarray}
L=-\frac{1}{2} \partial_\mu \phi \partial^\mu \phi - V(\phi)
\end{eqnarray}
with $V(\phi)=\lambda/4 (\phi^2-v^2)^2$ where $\lambda$ is the self-interaction coupling of the Higgs boson.  The Higgs boson's mass is given by $m=\sqrt{\lambda} v$. In an expanding universe, we can write the Higgs doublet as ${\cal H}=1/\sqrt{2}(0, \phi(\vec x,t)+\bar \phi(t)+v)^\top$, where $\bar \phi(t)$ is the cosmological background value of the Higgs field.
In the FLRW metric, and considering spatially homogeneous configurations, the equation of motion for the  time dependent  background Higgs field is given by
\begin{eqnarray}
\ddot{\bar \phi} + 3 H(t) \dot{\bar \phi} +V^\prime (\bar \phi)=0,
\end{eqnarray}
where the dot represents a time derivative and the prime a derivative with respect to the field $\bar \phi$. We will now look at small background Higgs field values around the minimum of the potential and thus neglect the quadratic term in the Higgs boson potential. For $V=1/2 m^2 \bar \phi^2$, and in the limit $H \ll m$ and $\dot H/H \ll m$, one makes the Ansatz 
\begin{eqnarray}
\bar  \phi(t)= A(t) \sin( m t + \theta)
\end{eqnarray}
where  $\theta$ is a phase. This leads to the following differential equation
\begin{eqnarray}
\dot A(t) = -\frac{3}{2} H A(t)
\end{eqnarray}
for $A(t)$ assuming that $\ddot A(t)$ is small.
It is straightforward to solve this differential equation in the limit where $H$ is constant (note that this is a good approximation in our current universe as the energy content is dominated by a cosmological constant), one finds
\begin{eqnarray}
A(t)=c e^{-\frac{3}{2} H t},
\end{eqnarray}
where $c$ is an integration constant.  We thus have
\begin{eqnarray}
\bar \phi(t)= c e^{-\frac{3}{2} H t} \sin( m t + \theta).
\end{eqnarray}
 This equation describes the cosmological time evolution of the Higgs field in an expanding universe with a nearly constant Hubble parameter. Note that the evolution of a scalar field in an expanding universe has been discussed before in different contexts, e.g., that of axions or dark energy \cite{Masso:2005zg,Preskill:1982cy,Abbott:1982af,Dine:1982ah} (see also \cite{Turner:1983he}).

Unless the parameter $c$ is fine-tuned to vanish, which we will argue is very unlikely given our knowledge of the early universe, the Higgs field will have a cosmological evolution. This cosmological evolution can be shifted in a cosmological time dependent vacuum expectation value:
\begin{eqnarray}
v(t)=v_0+c e^{-\frac{3}{2} H t} \sin( m t + \theta)
\end{eqnarray}
where $v_0=246$ GeV is the usual constant vacuum expectation value. We find
\begin{eqnarray}
    \dot v(t)= \frac{1}{2} c e^{\frac{-3 H t}{2}} \left ( 2 m \cos(m t+\theta) - 3 H \sin(m t+\theta) \right ),
\end{eqnarray}
where we treated $H$ as a constant.
In a year, the vacuum expectation value of the Higgs field will change by 
\begin{eqnarray}
\Delta v=c \left( e^{-\frac{3}{2} H t_1} \sin( m t_1 + \theta)- \sin(\theta) \right)
\end{eqnarray}
where $t_1=1$yr$ =3600 \times 24 \times 365/(6.582 \times 10^{-25}) \mbox{GeV}^{-1}=4.8 \times 10^{31} \mbox{GeV}^{-1}$. Today $H \sim 10^{-42}$ GeV, so in one year the vacuum expectation value changes by $\Delta v=(0.8\cos(\theta) - 1.6\sin(\theta)) c$.  Clearly this is potentially a large effect. Note that because of the large Higgs boson mass, 125 GeV, the oscillation frequency is extremely rapid and a priori difficult to measure.

Precise atomic clock measurements of the ratio $\mu=m_e/m_{p}$ enable us to derive a bound on $c$ and the phase $\theta$. A cosmological evolution of the vacuum expectation value of the Higgs boson will lead to a cosmological time dependence of  the masses of all the fermions and electroweak bosons. The mass of the electron $m_e$ is given by $m_e=\frac{1}{\sqrt{2}}\lambda_e v $ where $\lambda_e$ is the Yukawa coupling of the electron. The proton mass $m_p$ is however mainly determined by the scale of the strong interactions, $\Lambda_{QCD}$, the contributions of quark masses is very suppressed. Within the standard model this scale will not change with time. We can thus use the ratio of the mass of the electron to the mass of the proton to bound $c$ and $\theta$. We find that the change of $\mu$ in a year is given by 
\begin{eqnarray}
\frac{\Delta \mu}{\mu}\bigg |_{1 yr}= \int_0^{1 yr} \frac{\dot v(t)}{v_0} dt = \frac{c}{v_0}    \left( e^{-\frac{3}{2} H t_1} \sin( m t_1 + \theta)- \sin(\theta) \right)=
\frac{c}{v_0} (0.8\cos(\theta) - 1.6\sin(\theta)). \nonumber \\ 
\end{eqnarray}
The most stringent bound on $\Delta \mu/\mu$ comes from the comparison of the transitions in Yb$^+$ with the cesium atomic clock: $\Delta \mu/\mu=(-0.5 \pm 1.6) \times 10^{-16} \mbox{yr}^{-1}$, see e.g. \cite{Calmet:2014qxa} for a recent review. One obtains the following bound:
\begin{eqnarray}
c (0.8\cos(\theta) - 1.6\sin(\theta))<10^{-14} \ \mbox{GeV}.
\end{eqnarray}

From a theoretical point of view, we have little information on the value of $c$ and $\theta$. It is an initial condition problem, $c$ and $\theta$ are fixed by the initial value of the Higgs field at the start of our universe. However, as mentioned previously, it is very difficult to imagine that the Higgs field could start and remain at zero GeV during the course of our universe. Indeed during inflation, the Higgs field is essentially massless and it will thus be excited by the rapid expansion of our universe \cite{Vilenkin:1982wt,Vilenkin:1983xp,Linde:1982uu,Starobinsky:1982ee,Dolgov:2005se}. Given the apparent metastability of the electroweak vacuum, this is actually an issue as the Higgs field could easily fly over the $\sim10^9$ GeV barrier of the false electroweak vacuum and end up in the real vacuum with catastrophic consequences for the universe. This issue has led several authors to study mechanisms \cite{Lebedev:2012sy,Calmet:2017hja} that could lead the inflaton to drive the Higgs field to its false electroweak vacuum. In such scenarios, the Higgs field value can decay from a tenth of the Planck mass to the electroweak scale in about 20 e-foldings. Such models require a large total number of e-foldings, at least 100, which would lead to a $c\sim \exp(-3/2 H t)=7 \times 10^{-49}$GeV. Clearly, the value of $c$ is strongly model dependent: it depends on the specific inflation model one chooses, the initial condition for the Higgs field and the mechanism that drives the Higgs field to the false electroweak vacuum. In that sense measurements of $c$ and $\theta$ will enable us to probe early universe physics and a measurement with atomic clocks of these parameters could help to differentiate between different models of inflation or re-heating.

While the magnitudes of the parameters $c$ and $\theta$ which determine the cosmological time dependence of the Higgs boson's vacuum expectation value are strongly model dependent, a cosmological time evolution of the vacuum expectation value of the Higgs boson seems unavoidable given our current understanding of the standard models of particle physics and cosmology. 
Measurements of $c$ and $\theta$ would, within a specific model of inflation, enable us to determine the initial value of the Higgs field at the start of our universe. It would also be a test of the mass generation mechanism for fermions which would be complementary to the studies performed at the Large Hadron Collider at CERN. Remarkably, atomic clock experiments have the potential to not only  probe  fundamental high energy physics but also very early universe physics.

{\it Acknowledgments:}
This work is supported in part  by the Science and Technology Facilities Council (grant number  ST/J000477/1).


\bigskip{}


\baselineskip=1.6pt 

\begin{thebibliography}{10}


  
  \bibitem{Dirac} P. \ M. \ Dirac, Nature {\bf192}, 235 (1937).
  
 \bibitem{Milne} E. \ A. \ Milne, Relativity, Gravitation and World Structure,
Clarendon press, Oxford, (1935), Proc. Roy. Soc. A, 3, 242 (1937).
\bibitem{Jordan} P. \ Jordan, Naturwiss., 25, 513 (1937), Z. Physik, 113, 660 (1939).
  
\bibitem{Marciano:1983wy} 
  W.~J.~Marciano,
  Phys.\ Rev.\ Lett.\  {\bf 52}, 489 (1984).
  doi:10.1103/PhysRevLett.52.489
  
\bibitem{Uzan:2002vq} 
  J.~P.~Uzan,
  Rev.\ Mod.\ Phys.\  {\bf 75}, 403 (2003)
  doi:10.1103/RevModPhys.75.403
  [hep-ph/0205340].
  
\bibitem{Calmet:2001nu} 
  X.~Calmet and H.~Fritzsch,
  Eur.\ Phys.\ J.\ C {\bf 24}, 639 (2002)
  doi:10.1007/s10052-002-0976-0
  [hep-ph/0112110].
  
\bibitem{Calmet:2002ja} 
  X.~Calmet and H.~Fritzsch,
  Phys.\ Lett.\ B {\bf 540}, 173 (2002)
  doi:10.1016/S0370-2693(02)02147-0
  [hep-ph/0204258].
  
  
\bibitem{Calmet:2006sc} 
  X.~Calmet and H.~Fritzsch,
  Europhys.\ Lett.\  {\bf 76}, 1064 (2006)
  doi:10.1209/epl/i2006-10393-0
  [astro-ph/0605232].
  
  
\bibitem{Passarino:2001wx} 
  G.~Passarino,
  hep-ph/0108254.
  
\bibitem{Casadio:2007ip} 
  R.~Casadio, P.~L.~Iafelice and G.~P.~Vacca,
  Nucl.\ Phys.\ B {\bf 783}, 1 (2007)
  doi:10.1016/j.nuclphysb.2007.05.015
  [hep-th/0702175 [HEP-TH]].
  
\bibitem{Yoo:2002vw} 
  J.~Yoo and R.~J.~Scherrer,
  Phys.\ Rev.\ D {\bf 67}, 043517 (2003)
  doi:10.1103/PhysRevD.67.043517
  [astro-ph/0211545].
  
\bibitem{Dent:2001ga} 
  T.~Dent and M.~Fairbairn,
  Nucl.\ Phys.\ B {\bf 653}, 256 (2003)
  doi:10.1016/S0550-3213(03)00043-9
  [hep-ph/0112279].
  
\bibitem{Calmet:2014qxa} 
  X.~Calmet and M.~Keller,
  Mod.\ Phys.\ Lett.\ A {\bf 30}, no. 22, 1540028 (2015)
  doi:10.1142/S0217732315400283
  [arXiv:1410.2765 [gr-qc]].
  
\bibitem{Fritzsch:2016ewd} 
  H.~Fritzsch, R.~C.~Nunes and J.~Sola,
  Eur.\ Phys.\ J.\ C {\bf 77}, no. 3, 193 (2017)
  doi:10.1140/epjc/s10052-017-4714-z
  [arXiv:1605.06104 [hep-ph]].
  
\bibitem{Sola:2015xga} 
  J.~Solà,
  Mod.\ Phys.\ Lett.\ A {\bf 30}, no. 22, 1502004 (2015)
  doi:10.1142/S0217732315020046
  [arXiv:1507.02229 [hep-ph]].
  
\bibitem{Webb:2000mn} 
  J.~K.~Webb, M.~T.~Murphy, V.~V.~Flambaum, V.~A.~Dzuba, J.~D.~Barrow, C.~W.~Churchill, J.~X.~Prochaska and A.~M.~Wolfe,
  Phys.\ Rev.\ Lett.\  {\bf 87}, 091301 (2001)
  doi:10.1103/PhysRevLett.87.091301
  [astro-ph/0012539].
  
  
\bibitem{Masso:2005zg} 
  E.~Masso, F.~Rota and G.~Zsembinszki,
  Phys.\ Rev.\ D {\bf 72}, 084007 (2005)
  doi:10.1103/PhysRevD.72.084007
  [astro-ph/0501381].
  
\bibitem{Preskill:1982cy} 
  J.~Preskill, M.~B.~Wise and F.~Wilczek,
  Phys.\ Lett.\  {\bf 120B}, 127 (1983).
  doi:10.1016/0370-2693(83)90637-8
  
\bibitem{Abbott:1982af} 
  L.~F.~Abbott and P.~Sikivie,
  Phys.\ Lett.\  {\bf 120B}, 133 (1983).
  doi:10.1016/0370-2693(83)90638-X
  
\bibitem{Dine:1982ah} 
  M.~Dine and W.~Fischler,
  Phys.\ Lett.\  {\bf 120B}, 137 (1983).
  doi:10.1016/0370-2693(83)90639-1
  
\bibitem{Turner:1983he} 
  M.~S.~Turner,
  Phys.\ Rev.\ D {\bf 28}, 1243 (1983).
  doi:10.1103/PhysRevD.28.1243

  
  
\bibitem{Vilenkin:1982wt} 
  A.~Vilenkin and L.~H.~Ford,
  Phys.\ Rev.\ D {\bf 26}, 1231 (1982).
  doi:10.1103/PhysRevD.26.1231
  
\bibitem{Vilenkin:1983xp} 
  A.~Vilenkin,
  Nucl.\ Phys.\ B {\bf 226}, 527 (1983).
  doi:10.1016/0550-3213(83)90208-0
  
\bibitem{Linde:1982uu} 
  A.~D.~Linde,
  Phys.\ Lett.\  {\bf 116B}, 335 (1982).
  doi:10.1016/0370-2693(82)90293-3
  
\bibitem{Starobinsky:1982ee} 
  A.~A.~Starobinsky,
  Phys.\ Lett.\  {\bf 117B}, 175 (1982).
  doi:10.1016/0370-2693(82)90541-X
  
\bibitem{Dolgov:2005se} 
  A.~Dolgov and D.~N.~Pelliccia,
  Nucl.\ Phys.\ B {\bf 734}, 208 (2006)
  doi:10.1016/j.nuclphysb.2005.12.002
  [hep-th/0502197].

\bibitem{Lebedev:2012sy} 
  O.~Lebedev and A.~Westphal,
  Phys.\ Lett.\ B {\bf 719}, 415 (2013)
  doi:10.1016/j.physletb.2012.12.069
  [arXiv:1210.6987 [hep-ph]].
  
\bibitem{Calmet:2017hja} 
  X.~Calmet, I.~Kuntz and I.~G.~Moss,
  arXiv:1701.02140 [hep-ph].
  
\end{thebibliography}
\end{document}